\documentclass[global,twocolumn,hyperref]{svjour-arXiv}

\usepackage{graphics}
\usepackage{here}
\usepackage{textgreek}
\usepackage{cite}

\begin{document}
\title{Electric Field Effect on Optical Second-harmonic Generation 
of Amphoteric Megamolecule Aggregates\\
}
\author{\and{Yue Zhao}\inst{1}\hbox{\href{http://orcid.org/0000-0002-8550-2020}{\includegraphics{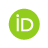}}} 
\and Yanrong Li\inst{1} \and Khuat Thi Thu Hien\inst{1} 
\and Goro Mizutani\inst{1}\thanks{\emph{Goro Mizutani:} mizutani@jaist.ac.jp}\hbox{\href{http://orcid.org/0000-0002-4534-9359}{\includegraphics{orcid.eps}}}
\and Nobuaki Ito\inst{2}
\and Harvey N. Rutt\inst{3}
\and Maiko Okajima\inst{1}
\and Tatsuo Kaneko\inst{1}
}                     
%
%

\institute{School of Materials Science, Japan Advanced Institute of Science and Technology, Asahidai 1-1 Nomi, 923-1292, Japan \and Center for Nano Materials and Technology, Japan Advanced Institute of Science and Technology, 1-1 Nomi, 923-1292, Japan \and School of Electronic and Computer Science, University of Southampton, SO17 1BJ, UK}

\date{
\\
\\
{\it J. Phys. Soc. Jpn.}, {\bf 86,} 124401 (2017).  DOI: 10.7566/JPSJ.86.124401}

\maketitle
\begin{abstract}
\bf We have detected second-order nonlinear optical response from an amphoteric mega-saccharides named sacran under a needle/ring electrode stimulus by using an optical second harmonic generation (SHG) microscope.  SHG was observed from sacran in the neighborhood of the negatively biased needle electrode below -4.5 V with respect to the outer ring electrode.  From the incident light polarization dependence of SHG, this sacran was suggested to be oriented toward the electrode needle.  SHG of the sacran polymer was judged to be induced by sacran aggregates with net positive charge.
\end{abstract}
\section{Introduction\label{Introduction}}

Algae-derived polysaccharides classified as red, brown, green, and blue-green algae are important components of life organism as extracellular polymeric substances (EPSs).  In particular, sulfated polysaccharides are abundantly present in nature, as typified by red alga-derived carrageenan { \cite{p,dm3}}, and show important vital functions such as induction of bone formation, adipogenesis and chondrogenic differentiation in stem cells { \cite{s}}.  In addition, the sulfated polysaccharides by chemical treatment can induce some important physiological function such as anti-virus and anti-cancer activities { \cite{uryu1991,uryu1992,uryu1999,uryu2003}}.  Sulfated polysaccharides are also widely attracting attention in the fields of food, cosmetics, pharmacy, medicine, and so on { \cite{md}}.

A blue-green alga is considered as the origin of plant chloroplast.  It has been reported that it can produce EPSs such as sulfated polysaccharides { \cite{k4}}.  However, many of blue-green algae are not suited to mass farming, so the study of its EPSs is basically blank in materials science.  Sacran, the target of this research, is a sulfated natural polysaccharide efficiently produced by EPSs of  {\it Aphanothece sacrum} { \cite{1}}.  {\it Aphanothece sacrum} is a rare blue-green algae, and its large-scale culturing method has been established.  Sacran is a non-crystalline supergiant polysaccharide with an absolute molecular weight of 10-30 Mg/mol { \cite{2}}.  Since the sulfation rate is about 11\% per sugar residue, a sacran molecular chain has 10,000 sulfate groups, and this number is unprecedented { \cite{1}}.  Therefore, it is considered to have very high anti-inflammatory and anti-allergic activities { \cite{ngatu2015sacran2,ngatu2015sacran1,aaai}}.  It has excellent substance sorptivity such as the efficient adsorption of metal cations { \cite{aaai,5,4,3,7,9}} and water-retention capacities of 6100 times its dry weight for pure water and 2600 times for saline { \cite{2}}.  

Since sacran has a small amount of amino groups of 1\% or less, it will form a various self-assembled structures in solution according to the environmental changes of concentration of sacran and the surrounding ions.  Therefore, sacran can also be an amphoteric electrolyte sulfated polysaccharide (ca. 100,000 anions and 1,000 cations in a chain { \cite{2}}).  For example, when the concentration reaches a certain value, the sacran molecules will change from a spherical to rod-like organization { \cite{10}}.  The rod-like molecules of sacran is also autonomously oriented to form a liquid crystal state { \cite{5,10,39,polymer99}}.  In addition, sacran have a high content of anionic sugars with carboxylic acid (22 mol\%) and sulfate groups (11 mol\%) and a low content of cationic amino sugars.  It is an ampholytic sugar chain with an imbalanced charge ratio { \cite{1,2}}.  In our previous studies, we observed SHG images from the macroscopic non-centrosymmetric orientation part of the high purity sacran aggregates, but the origin of the anisotropic structure orientation in the sacran molecular aggregates was still unexplained { \cite{Zhao_sacran_JOSAA_2017}}. 

Optical second harmonic generation (SHG) used in this study is a second-order nonlinear optical phenomenon.  When light pulses pass through noncentrosymmetric materials such as chiral molecules or noncentrosymmetric structural units, they generate frequency doubled pulses.  It is well known that noncentrosymmetric crystals have non-zero second-order nonlinear susceptibilities { \cite{franken1961generation,bloembergen1962light,BNO}} and some ordered structures enhance the second-order nonlinear susceptibility { \cite{14,g0-12,13,32,Zhao_2017_spider_silk_SHG}}.  Sacran aggregates can generate SHG light because of their macroscopic ordered structure { \cite{Zhao_sacran_JOSAA_2017}}.  Zhao et al { {\cite{Zhao_sacran_JOSAA_2017}}} reported that SHG microspots of several tens of \textmu m size were observed from cotton-like lump, fibers, and cast films of sacran.  The polarization-dependent SHG microscopic images showed that SHG spot has a concentric multilayer structure of liquid crystal domains.  Each domain structure has its own orientation.  Rather spatially continuous SHG was also observed near the edges of the cast film made from sacran aqueous solution.  One of the candidate origins of this more continuous SHG is a nonuniform concentration distribution of sacran in the films caused by the different evaporation velocity of water from the solution droplet on the substrate at the stage of sample preparation.  On the other hand, the SHG microspots disappeared after filteration of a source aqueous solution with a pore size of 0.45 \textmu m.  The origins of these SHG microspots was still not perfectly clear.

In this study, in order to get a hint of the origins of the SHG of sacran observed so far, the anions and cations of the sacran having various structures were attempted to be separated by electrophoresis.  Anionic and cationic aggregates of sacran are expected to gather around the negative and positive electrodes, respectively, and then the SHG images of such films near the electrodes were obtained for the first time.  In particular, the electric field near the center needle electrode is extremely large, and hence the charged sacran molecules were expected to be attracted and efficiently oriented.  Sacran has a large amount of strong electrolytes of the sulfate group.  The accumulation state of functional groups in a sacran molecular film was also examined by tracking specific elements such as sulfur and nitrogen by X-ray photoelectron spectroscopy (XPS) measurement.

\section{Materials and Method \label{Materials and Method}}

\begin{figure}[h]
\centering
\begin{minipage}[t]{6.5cm}
\resizebox{1\textwidth}{!}{  \includegraphics{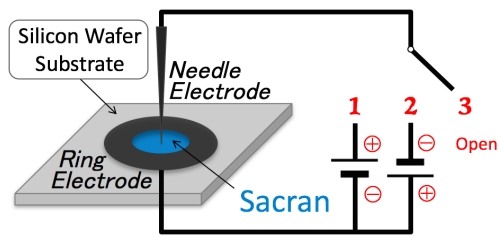}}
\end{minipage}%
\caption{Schematic electrode system used in this study. The needle/ring electrode was composed of a needle electrode and a ring electrode on a silicon wafer substrate.  A sacran aqueous solution was dropped inside the ring, and then the needle electrode was placed at the center.  Samples were made with the switch positions 1, 2 and 3.  The current was made to flow until the films became completely dry.}
\label{Electrode_setup}  
\end{figure}

The high purity sacran was used as received from Green Science Material Inc (Kumamoto, Japan).  Samples in the electrode setup as shown in Fig. \ref{Electrode_setup} were made with a sacran aqueous solution with concentration of 0.5\%.  The photograph of the electrode setup is shown in Fig. S1 (Supplementary Information).   We also filtered the sacran aqueous solutions with concentration of 0.5\% by membrane filters of 1.2 \textmu m and 0.45 \textmu m pore sizes.  Samples were made with the switch positions 1, 2 and 3 in Fig. \ref{Electrode_setup} using the unfiltered solution and filtrate at the two pore sizes.  The current was made to flow until the films became completely dry.  The total number of samples were nine.  We carried out the experiment with steel or gold electrodes and compared the result in order to check the electrode material influence on the result.  When the outer ring electrode was made by winding a gold wire on a plastic ring, a gold needle was used as the counter electrode.  When a steel ring was used as the outer electrode, a steel needle was used as the counter electrode.  The diameter of the needle was about 50 \textmu m, the inner diameter of the other electrode ring was about 6.6 mm, and the thickness of the ring was 1.0 mm.  The needles lightly touched the silicon substrates.  The ring electrodes were pressed against the silicon substrates by strong alligator clips.  The position of the tip was within 0.2 mm from the center of the ring.  The surface of the silicon wafer appeared to have thick native oxide layers on their surfaces and their resistance measured by an electric checker was more than 5 M\textOmega.  In the apparatus shown in Fig. \ref{Electrode_setup} and Fig. S1 (Supplementary Information), the electric current before dropping sacran aqueous solution into the cell was less than 10 \textmu A.  The observation method for observing SHG was the same as that described in our previous report { \cite{Zhao_sacran_JOSAA_2017}}.

\section{Result\label{Result}}

\subsection{Current change with time\label{Current change with time}}

\begin{figure}[h]
\begin{minipage}[t]{8.5cm}
\resizebox{1\textwidth}{!}{  \includegraphics{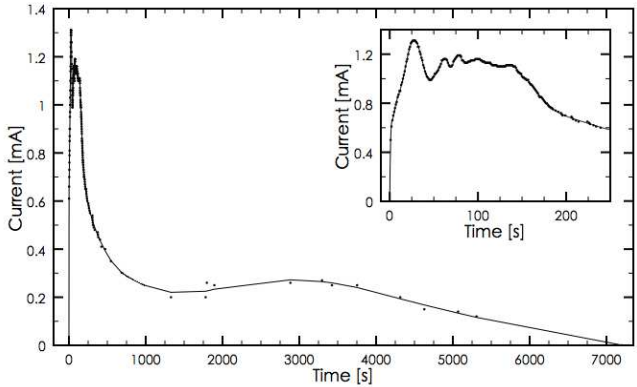}}
\end{minipage}%
\caption{Current as a function of time through the sacran aqueous solution when the applied voltage was -6 V.  At time zero the current was switched on and at time 7000 seconds the sample was dry.  The inset shows the chart with the time scale expanded.}
\label{I_t_curve}  
\end{figure}

Figure \ref{I_t_curve} shows the current as a function of time after 3 drops (about 70 \textmu L) of high purity sacran aqueous solution with the concentration of 0.5\% was put between the needle electrode biased at -6 V and the ring electrode.  The switch of the power supply was turned on immediately after the needle electrode was placed.  Only when the switch position was 2 in Fig. \ref{Electrode_setup}, gas bubbles were generated around the negatively biased needle and the positively biased ring electrodes during the first few minutes.  In order to check whether the sacran was deteriorated or not by the electric current, we took the resultant material from the cell and put it into hot water at temperature of 90$^\circ$C to 100$^\circ$C.  We found that the material partly solved and partly gelled and this fact showed that the sacran did not experience a remarkable chemical change by the electric current.  The aqueous solution dried slowly until it was completely dry after about 2 hours.  The electric current was zero when the film was dry.  The drying speed was not affected by the applied voltage.  In Fig. \ref{I_t_curve} an exponential-like decay of current is seen up to 1200 seconds and a low broad peak is seen at 3000 seconds.  In the current profile in the initial 4 minutes in the inset one sees repeated up and down. 

\subsection{SHG of sacran for different applied voltages\label{SHG of sacran for different applied voltages}}

\begin{figure*}[h]
\centering
\begin{minipage}[t]{12.5cm}
\resizebox{1\textwidth}{!}{  \includegraphics{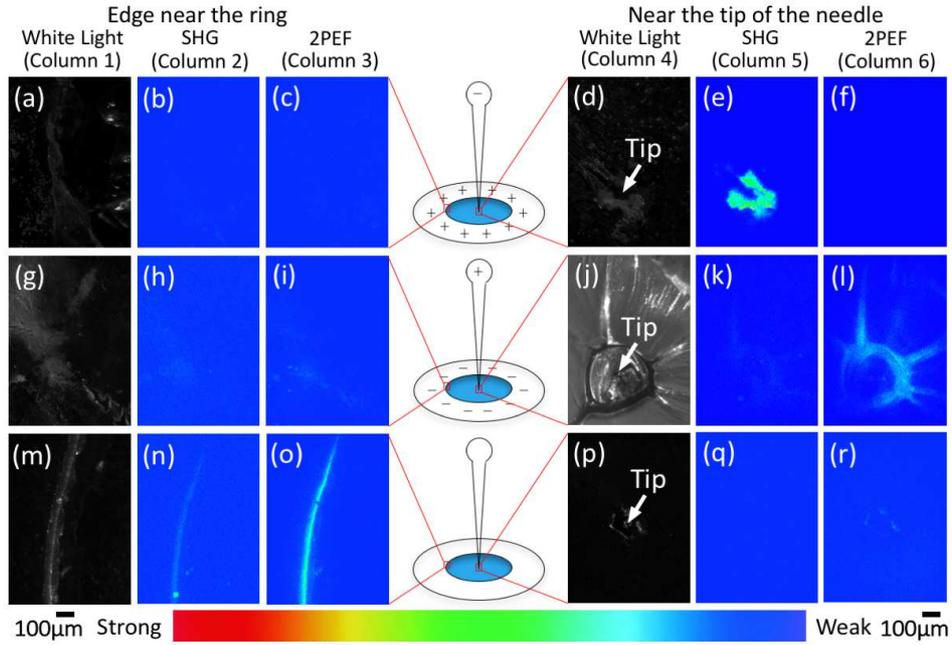}}
\end{minipage}%
\caption{The nine images on the left are the images of the sacran film near the inner edge of the steel ring electrode.  The nine images on the right are the images near the central steel needle electrode.  For (a-f) the needle electrode was negatively biased, for (g-l) the needle electrode was positively biased, and for (m-r) voltage was not applied during the film formation.  In (a, g, m, d, j, and p) microscopic images using white light illumination and a CMOS camera (Manufacturer: Lumenera corporation, model number: Lu135M) are shown.  In (b, h, n, e, k, and q) SHG images observed by using a bandpass filter of 400 nm wavelength, and in (c, i, o, f, l, and r) 2PEF images observed by using a bandpass filter of 438 nm wavelength are shown.  The magnification of the objective lens was $\times$5 (NA= 0.15).  For SHG and 2PEF images the incident light wavelength was 800 nm, the area of the beam on the sample was 6 mm$^2$ and the excitation power was 20.1 mW.  The integration time of the imaging was 60 s.  The diameter of the tip of the needle was about 50 \textmu m, and its position is indicated by white arrows in (d), (j), and (p).}
\label{SHG_by_electrode}
\end{figure*}

\begin{table*}[h]
\caption{Summary of observation results of SHG near each electrode}
\centering
\label{SHG_each_electrode}
\begin{tabular}{ccccccccccccc}
\cline{1-13} 
\multicolumn{1}{c}{} & \multicolumn{3}{c}{Near needle} & \multicolumn{1}{c}{} & \multicolumn{3}{c}{Near ring} & \multicolumn{1}{c}{} & \multicolumn{4}{c}{Voltage {[}V{]}} \\ \cline{2-4} \cline{6-8} \cline{10-13} 
                      & \,\, + \,\, & \,\, - \,\, & \,\, 0 \,\, & \,\,\,\, & \,\, + \,\, & \,\, - \,\, & \,\, 0 \,\, & \,\,\,\, & \, 1.5 \, & \, 3.0 \, & \, 4.5 \, & \, 6.0 \, \\
\cline{1-13}
\,\, SHG \,\, &  No  & Yes       & No       &                       & No       & No       & Yes      &                       & No     & No     & Yes     & Yes      \\
2PEF      & Yes       & No        & No       &                       & No       & No       & Yes      &                       & -       & -       & -      & - \\ 
\hline
\end{tabular}
\end{table*}

\begin{figure}[h]
\centering
\begin{minipage}[t]{7.5cm}
\resizebox{1\textwidth}{!}{  \includegraphics{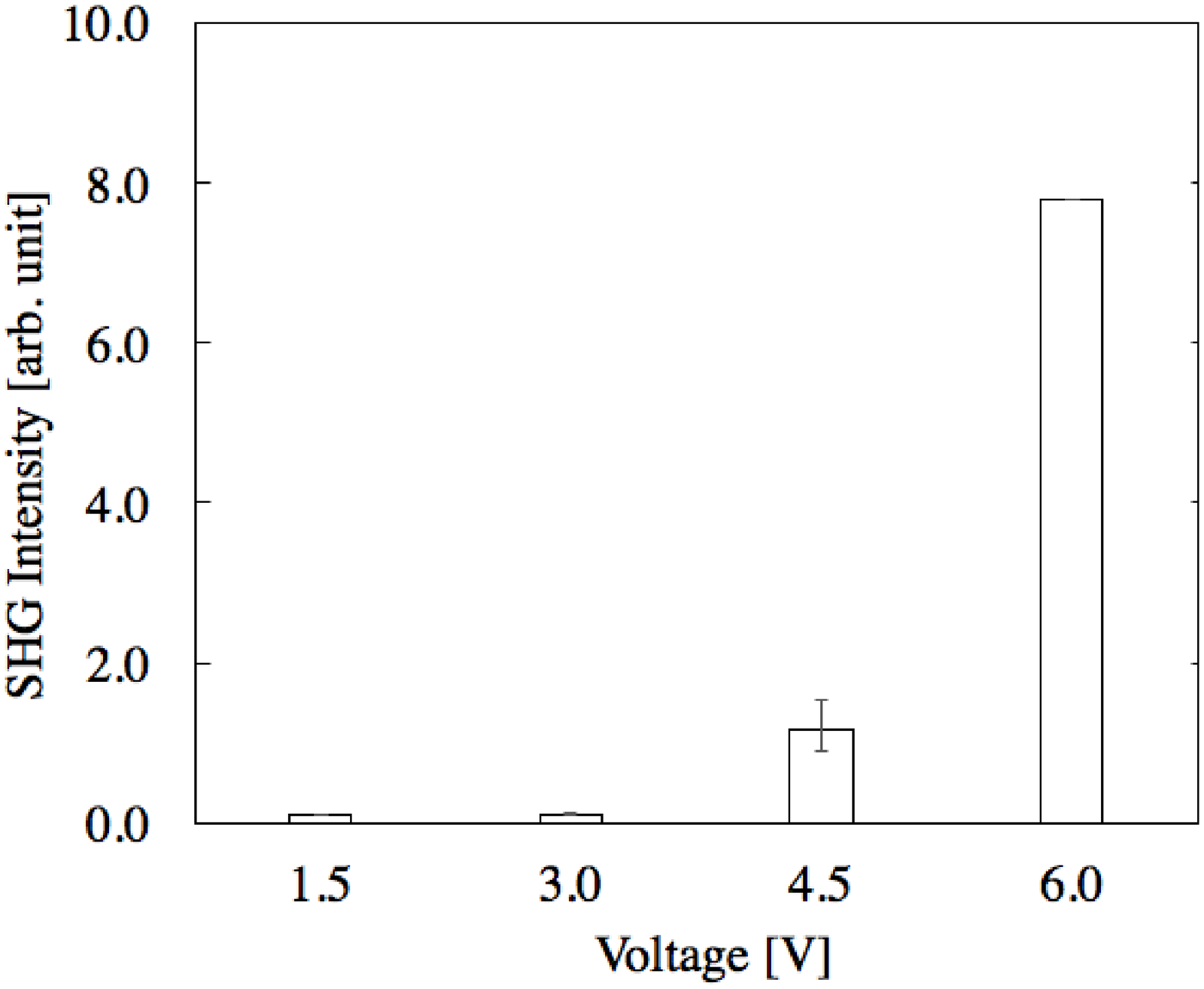}}
\end{minipage}%
 \caption{Average intensity of SHG signal around the negative needle electrode as a function of voltage.}
\label{Voltage_SHG}
\end{figure}

In Fig. \ref{SHG_by_electrode}, the nine images on the left are images near the inner edge of the steel ring electrode.  The nine images on the right are images near the tip of the central steel needle electrode.  In Fig. \ref{SHG_by_electrode} (a-f) the needle electrode was negatively biased, in (g-l) the needle electrode was positively biased, and in (m-r) no voltage was applied during the film formation.  The images in the same column are either microscopic images of white light illumination, SHG, or 2-photon excitation fluorescence (2PEF), as they are shown at the top of the figure.  As it can be seen from Fig. \ref{SHG_by_electrode}, when a voltage of -6 V was applied to the needle electrode with respect to the outer ring, a strong SHG signal was observed around the negative needle electrode (Fig. \ref{SHG_by_electrode} (e)).  On the contrary, there was no SHG signal observed near the positive ring electrode (Fig. \ref{SHG_by_electrode} (b)).  For the voltage in the reverse direction, there was no SHG signal around the positive needle electrode (Fig. \ref{SHG_by_electrode} (k)), and the SHG signal near the negative ring electrode was also negligible (Fig. \ref{SHG_by_electrode} (h)).  When no voltage was applied, an SHG spot (Fig. \ref{SHG_by_electrode} (n)) of several tens of micrometer size was observed on the inner edge of the ring, and no SHG signal was observed around the needle (Fig. \ref{SHG_by_electrode}(q)).  On the other hand, a strong 2PEF signal was observed around the positive needle electrode (Fig. \ref{SHG_by_electrode} (l)).  There was no 2PEF signal observed near the negative needle electrode (Fig. \ref{SHG_by_electrode} (f)).  These results are summarized in Table \ref{SHG_each_electrode}.

The dependency of the SHG signal near the negative needle electrode on the applied voltage was also checked.  For 4.5 V the SHG signal was detectable around the negative needle electrode, while the SHG disappeared for the voltage of 3.0 V and 1.5 V.  This result is summarized in Fig. \ref{Voltage_SHG}.  

In order to check the influence of the electrode material, we changed the material of the both electrodes to gold, made three kinds of samples in the same way as in Fig. \ref{Electrode_setup}, and observed their SHG images.  As a result, SHG signal was observed around the negative gold needle electrode just like the case with the steel electrode.  Therefore, we can say that the results do not depend on the material of the electrodes.  

In the case of the samples made with the filtrate at 1.2 \textmu m and 0.45 \textmu m pore sizes, SHG signal was observed and no 2PEF signal was observed near the negative needle electrodes for all the samples.  No SHG signal was observed at the portion in the middle of the needle and the ring in any sample.

\subsection{Incident polarization dependence of SHG\label{Incident polarization dependence of SHG}}

\begin{figure}[h]
\centering
\begin{minipage}[t]{8.5cm}
\resizebox{1\textwidth}{!}{  \includegraphics{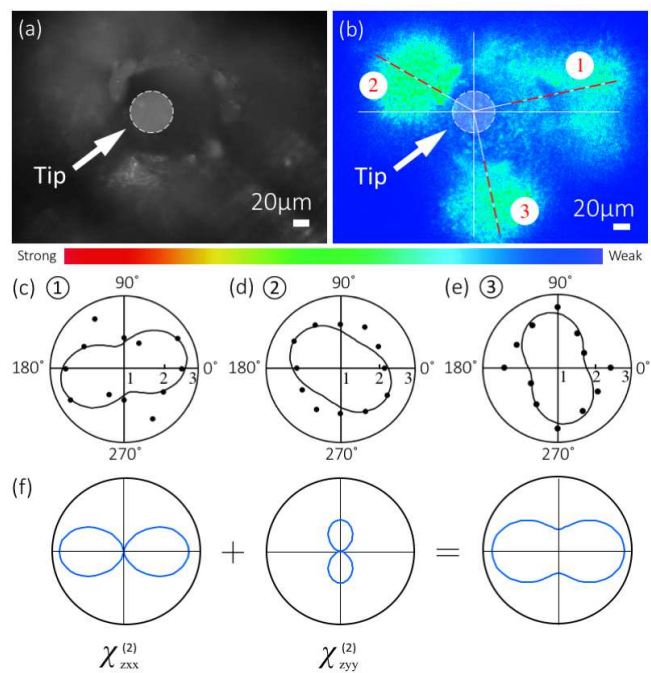}}
\end{minipage}%
 \caption{(a) A linear microscopic image of a sacran film around the negative needle electrode obtained with a white light illumination and a CMOS camera. (b) SHG image of the same film.  (c-e) Incident polarization dependence of average SHG intensities integrated on the three dashed lines. The plotted dots from 0$^\circ$-180$^\circ$ are the measured values, and those from 210$^\circ$-330$^\circ$ are copies of 30$^\circ$-150$^\circ$ values, since the polarizations at 210$^\circ$-330$^\circ$ are equivalent to 30$^\circ$-150$^\circ$.  The solid lines show calculated patterns of SHG intensity using the nonlinear susceptibility elements of the sacran film stated in the text.  (f) SHG polar graphs of two components of nonlinear susceptibilities in sacran and their sum.}
\label{Electrode_SHG_polarization}
\end{figure}

We investigated the incident polarization dependence of the strong SHG of the sacran around the negative needle electrode.  As shown in Fig. \ref{Electrode_SHG_polarization} (b), we have integrated the SHG intensity along the dashed lines in Fig. \ref{Electrode_SHG_polarization} (b) and obtained the dependence on the incident light polarization as shown in Fig. \ref{Electrode_SHG_polarization} (c, d, e).  Here, the scale of the SHG intensity axis starts from 1 in order to make the polar polarization dependence shapes clearer.  As shown in Fig. \ref{Electrode_SHG_polarization}(f), the assumed nonzero components of nonlinear susceptibility of the sacran film are ${\chi }^{\rm (2)}_{\rm zxx}$ and ${\chi }^{\rm (2)}_{\rm zyy}$ in the molecular coordinate (x, y, z). Here x is the molecular axis of sacran.  When the molecular axis rotates by angle $\phi$, the nonzero nonlinear susceptibility elements become
\begin{eqnarray}
&&{{\chi}^{\rm (2)}_{\rm zxx}}^{\prime}(\phi)={\chi}^{(2)}_{\rm zxx }{\rm cos}^{ 2 }\phi +{ \chi  }^{ (2) }_{\rm zyy}{\rm sin}^{ 2 }\phi \\
&&{{\chi}^{\rm (2)}_{\rm zxy}}^{\prime}(\phi)={ (-{\chi}^{ (2) }_{\rm zxx}+{ \chi  }^{ (2) }_{\rm zyy}})2{\rm sin}\phi {\rm cos}\phi \\
&&{{\chi}^{\rm (2)}_{\rm zyy}}^{\prime}(\phi)={\chi}^{(2)}_{\rm zxx}{\rm sin}^{2}\phi +{\chi}^{(2)}_{\rm zyy}{\rm cos}^{2}\phi .
\end{eqnarray}
The SHG patterns calculated with these ${{\chi}^{\rm (2)}_{\rm ijk}}^{\prime}(\phi)$ are shown by the solid lines in Fig. \ref{Electrode_SHG_polarization} (c, d, e).  From the $\phi$ values of each area we can infer that the sacran molecules are oriented at angles of $\phi_{1}=12^{\circ}$, $\phi_{1}=150^{\circ}$, $\phi_{1}=282^{\circ}$.

\subsection{Measurement results of X-ray photoelectron spectroscopy\label{Measurement results of X-ray photoelectron spectroscopy}}

In order to investigate the atomic elements contained at each position of the sacran film, we performed X-ray photoelectron spectroscopy (XPS) on the films created above by using S-Probe TM 2803 model manufactured by Fisons Instruments Co., Ltd., USA.  The results are shown in Table \ref{XPS_table} and Figs. S2 and S3 (Supplementary Information).  In the sacran aqueous solution, the sulfur atom tends to have a negative charge in a sulfate group, the sodium atom to be a counter cation, and the nitrogen atom in amino sugar to be a cation.  The negative charge of sacran anion mainly originate from carboxylic acid (22 mol\%) and sulfate groups (11 mol\%) { \cite{1,2}}.  The anion can be identified by tracking the sulfur.  Sodium was also a candidate origin of SHG activation, so we also monitored sodium signal.  Finally, nitrogen is also an important element in sacran although the amount is small, and its signal was also monitored.  Table \ref{XPS_table} shows that N is present everywhere in the sample.  This is a reasonable result, since sacran has amino groups containing nitrogen.  On the other hand, the sulfur was detected everywhere except around the negatively biased needle electrode.  Irrespective of the presence of the applied voltage, sodium was always absent near the film outer edge.  The detailed spectrum obtained by the XPS measurement for each element are shown in Figs. S2 and S3 (Supplementary Information).

\begin{table*}[h]
\centering
\caption{Measurement result of XPS}
\label{XPS_table}
\begin{tabular}{cccccc}
\hline
\multicolumn{1}{c}{} & \multicolumn{2}{c} \,\,\, {Without electrode} \,\,\, & \multicolumn{1}{c}{} & \multicolumn{2}{c}{With electrodes (ring:+, needle:-)}  \\ \cline{2-3} \cline{5-6} 
 & \,\, Center \,\, & Edge & \,\,\,\, & \,\,\,\,\,\, Center \,\,\,\,\,\, & Edge \\ \hline 
 \,\, N \,\, & \,\, Yes \,\, & \,\, Yes \,\, & \,\,\,\, & \,\, Yes \,\, & \,\, Yes \,\, \\
 \,\, S \,\, & \,\, Yes \,\, & \,\, Yes \,\, & \,\,\,\, & \,\, No \,\, & \,\, Yes \,\, \\
 \,\, Na \,\, & \,\, Yes \,\, & \,\, No \,\, & \,\,\,\, & \,\, Yes \,\, & \,\, No \,\, \\
\hline
\end{tabular}
\end{table*}

\section{Discussion\label{Discussion}}

Since sacran is an amphoteric electrolyte, it is electrolytically separated in an electric field, and the anion gathers around the positive electrode, and the cation gathers around the negative electrode.  As shown in Fig. \ref{I_t_curve}, the current rose up to $\sim$1.2 mA in 20 seconds and lasted for $\sim$240 seconds after applying the voltage, and its origin maybe mainly the electrolysis of water.  Some bubbles were generated strongly from the negative needle and the positive ring electrodes when the current was high in $\sim$240 seconds at the beginning, and then gradually stopped in next 300 seconds.  The current reduced to $\sim$0.2 mA and became constant from $\sim$1000 to 3500 seconds.  The constant current should be due to cationic and anionic motion of the amphoteric sacran molecules but its detailed origin is unclear.  After 3500 seconds, the sample became dry and the concentration of the solution became condensed and it resulted in gradual decrease of the current.  

The vibration of the ion current as a function of time seen in the inset of Fig. \ref{I_t_curve} can have two candidate origins.  As candidate origin (i): In the first few minutes, the gas bubbles generated from the needle may have blocked the electrical contact between the needle electrode and the solution, then the bubbles broke and the needle electrode and the solution contacted each other again.  Thus the contact area between the needle and the solution changed as a function of time.  As a result, in the first few minutes, the ion current shows the oscillation as shown in the inset of Fig. \ref{I_t_curve}.  As candidate origin (ii): There are substances with different viscosities and sizes.  The functions and relaxation times of the moving speed of each substance are different.  A curve like the inset of Fig. \ref{I_t_curve} may correspond to the distribution of the arrival time of those substances.  We guess that the gas bubbles were generated by the electrolysis of water.  So hydrogen gas was generated at the negative electrode and oxygen gas was generated at the positive electrode.  The reason why it only happened for the switch position 2 in Fig. \ref{Electrode_setup} is unknown.

As shown in Table \ref{SHG_each_electrode}, the SHG signal was observed from around the negatively biased needle electrode, while there is no SHG signal observed from around the positive electrode.  When we prepared the sample, the sacran film dried gradually under the applied voltage.  Then the applied voltage was removed and SHG was observed.  Therefore, the SHG was not the so-called Electric field induced SHG (EFSHG).  From these facts, it can be inferred that the origin of SHG is sacran cations.  One of the reasons for the strong SHG around the negative \textquoteleft needle\textquoteright \, electrode is the high-density sacran cation aggregating because of the extremely high electric flux density around the needle electrode.  Here we do not deny that sacran anions are also attracted to the cations to form a well-ordered domain.  On the other hand, at the inner edge of the ring electrode, the electric flux density was low, and the Lorentz force pulling the ions was weak.  Assuming that the same amount of molecules moved in the positive and negative biases, the molecular density near the ring electrode is much smaller than around the needle.  That is why SHG was generated more efficiently near the needle electrode.    

\begin{figure}[h]
\centering
\begin{minipage}[t]{5.5cm}
\resizebox{1\textwidth}{!}{  \includegraphics{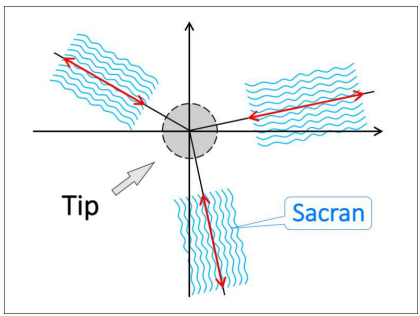}}
\end{minipage}%
\caption{A schematic model for sacran molecule orientation near the negative needle electrode. Sacran molecules (wavy lines) were oriented in the direction of the needle. The red arrows indicate the incident polarization direction of the strongest SHG response.}
\label{Electrode_SHG_polarization_discussion}
\end{figure}

Next, we consider the dependence of the SHG on the incident light polarization around the negative needle electrode.  There seems to be molecular orientation pointing towards the needle electrode as a schematic model shown in Fig. \ref{Electrode_SHG_polarization_discussion}.  Sacran molecular aggregates exhibit a macroscopic asymmetric structure, and the SHG was activated.

As shown in Fig. \ref{Electrode_SHG_polarization}(b), the SHG activation around the negative needle electrode is not uniform.  The reason for thzis is considered as follows.  Some hydrogen gas bubbles were generated from the negative needle electrode for the first few minutes after switching on.  After a few minutes, the free-flowing liquid almost disappeared and the bubbles disappeared.  Orientation could only occur in the first few minutes, and the orientation of sacran molecules under the gas bubbles was prevented because the molecular orientation is disturbed by the bubbles.

We must also check sodium ions as the candidate origin of the enhanced SHG.  When the sacran aqueous solution formed a film, the sodium ions moved to the center as the water evaporates.  This is considered to be the reason for the fact that sodium gathered to the center in the sacran film as can be confirmed in Table \ref{XPS_table}, regardless of the presence of the applied fields.  For any applied voltage, the SHG intensity and the concentrations of Na do not correlate with each other.  In addition, a sacran aqueous solution with a large amount of Na$^+$ dried without applying a voltage showed no SHG.  Therefore, sodium ions did not induce SHG in sacran.  The origins of SHG and fluorescence are different. SHG is due to the inversion symmetry breaking of the structure, while fluorescence is due to the electronic states of the substance.  The electrode polarity dependence of the fluorescence was observed in Fig. \ref{SHG_by_electrode} (l) and Fig. \ref{SHG_by_electrode} (f).  Therefore, the component with fluorescent property in sacran might have been separated, but the origin of the fluorescent property is unknown.  There is no fluorescence band at energy of 3.10 eV (400 nm), so far as literature is surveyed.  The fluorescence might be due to some imperfection of sacran or stimulated Raman scattering.

Since no sulfur was detected from the sacran aggregates near the negative needle electrode as shown in Table \ref{XPS_table}, the component with almost no sulfate group is suggested to have gathered there.  Natural sulfated polysaccharides have corresponding ingredients.  For example, carrageenan from red algae has many sulfate groups on the average, but, it has a component called \textbeta -carrageenan with no sulfate group { \cite{p105}}.  On the other hand, muramic acid, an important component in sacran, have nitrogen.  Sugar residues with muramic acid combine with sacran cations to become positively charged bodies.  The positively charged bodies may have gathered around the negatively biased needle electrode.  They may have been oriented and have activated the SHG.  Nitrogen was also detected near the positively biased ring electrode.  This nitrogen is also considered to be from muramic acid.  The weight percentage of nitrogen (0.30\%) in sacran is extremely small compared to sulfur (2.07\%) { \cite{1,2}}.  The muramic acid are bound to sugar residues together with large amounts of sulfate group and carboxylic acid and the net charge is negative.  Hence the sulfate group and carboxylic acid may have been moved toward the positive electrode. 

In our previous paper { \cite{Zhao_sacran_JOSAA_2017}}, we reported that SHG microspots were observed in cast films made from sacran aqueous solution and they disappeared by filtration with a 0.45 \textmu m pore size of the solution before casting.  We can raise two nonexclusive candidate origins for this effect considering the results of the current study.  In Candidate origin (1): the aggregates generating SHG themselves were removed by the filter.  In Candidate origin (2): the aggregates generating SHG were disintegrated into smaller molecules under the pressure of the filtration and then they lost their SHG activity by disintegration.  On the other hand, in this experiment, it was observed that SHG was generated by stimulation of the electric field, even if the sacran aqueous solution was filtered with the pore size of 0.45 \textmu m.  Right now which candidate is feasible is not known and it is our future research topic.

\section{Summary\label{Summary}}

Sacran films under electric field in needle/ring electrodes were fabricated, and the sacran molecules were found to generate SHG only near the negative electrode.  In particular, extremely strong SHG signals just around the negatively biased needle electrode were observed.  From these results, the origin of SHG of the sacran polymers is judged to be the sacran cations.  The incident polarization dependence of the SHG around the negative needle electrode strongly suggests that the sacran molecules were oriented toward the center of the needle electrode.  The XPS measurement told us that the sacran without sulfate groups mainly aggregates around the negatively biased electrode and can be related to SHG enhancement.  Thus, this result suggests one method for increasing the sulfate group content of sacran and confirming the resulting product.

\bibliographystyle{0.zidingyi}
\bibliography{sacran_ele_ref}

\newpage

\begin{figure*}[h]
{\bf Supplementary Information}\\

{\it 1. Electrode setup}\\

\centering
\begin{minipage}[t]{18cm}
\resizebox{1\textwidth}{!}{  \includegraphics{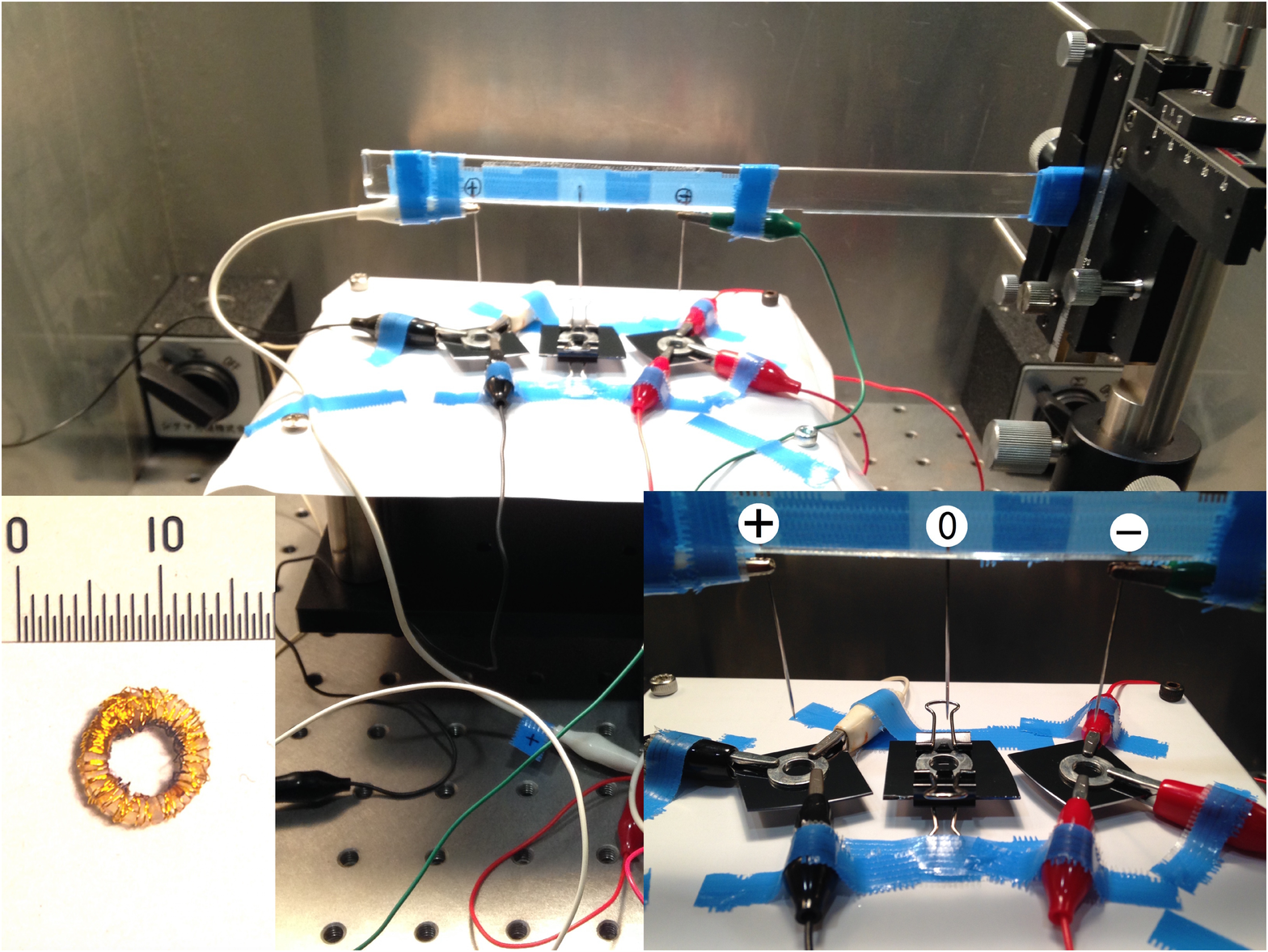}}
\end{minipage}\\%
\flushleft

{\bf Figure S1.}
Experimental circumstances: Electrode setup.
\label{Figure_S1}
\end{figure*}

\begin{figure*}[h]
{\it 2. Measurement result of X-ray photoelectron spectroscopy (XPS)}\\

\centering
\begin{minipage}[t]{18cm}
\resizebox{1\textwidth}{!}{  \includegraphics{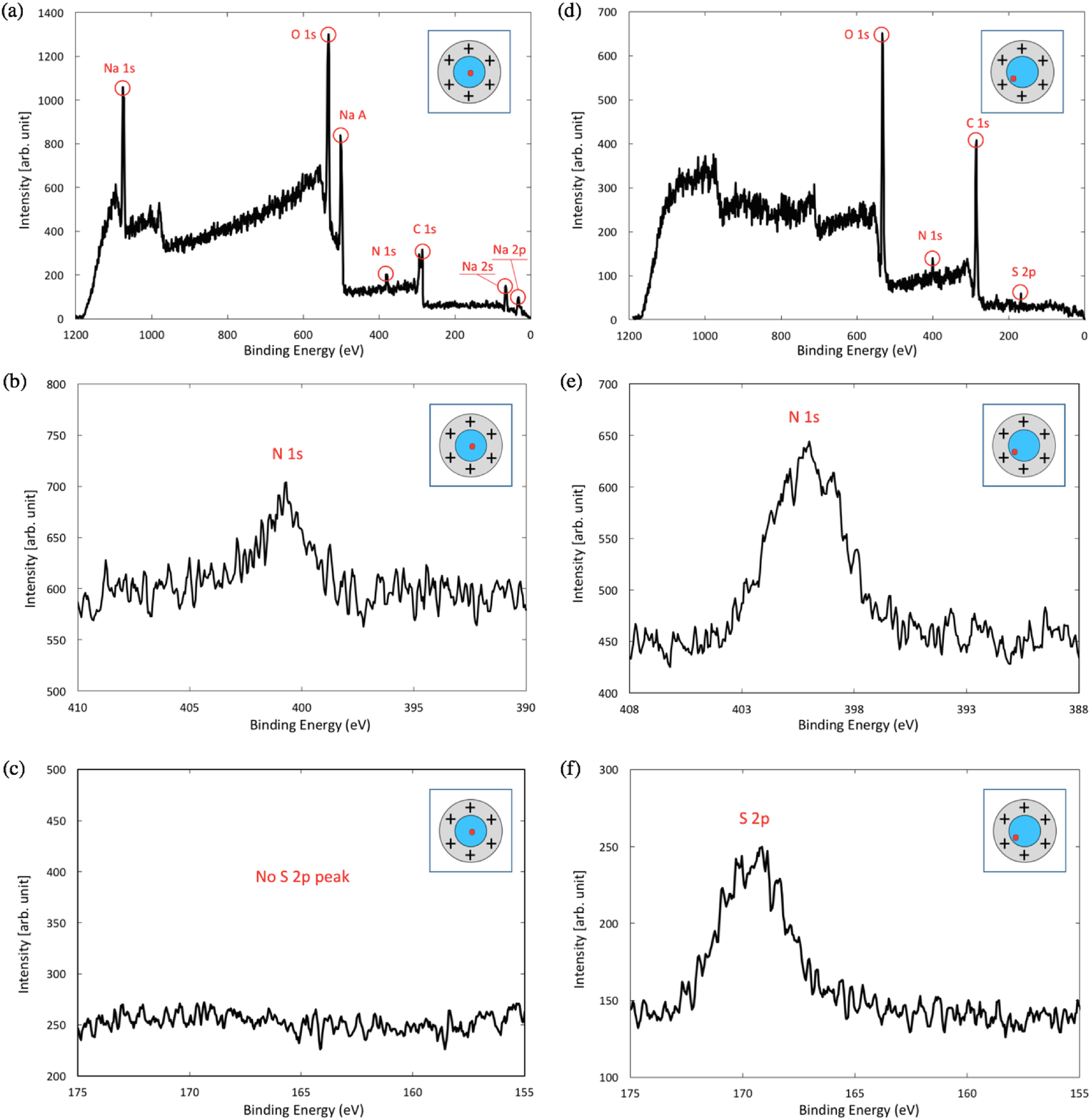}}
\end{minipage}\\%
\flushleft
{\bf Figure S2.} (a-c) XPS spectrum around the needle of the negative electrode.  (b) Result of narrow scan of the binding energy in the 1s orbital of N.  (c) Result of narrow scan of the binding energy in the 2p orbital of S.  (d-f) XPS spectrum near the ring of the positive electrode.  (e) Result of narrow scan of the binding energy in the 1s orbital of N.  (f) Result of narrow scan of the binding energy in the 2p orbital of S.
\label{Figure_S2}
\end{figure*}

\begin{figure*}[h]

\centering
\begin{minipage}[t]{18cm}
\resizebox{1\textwidth}{!}{  \includegraphics{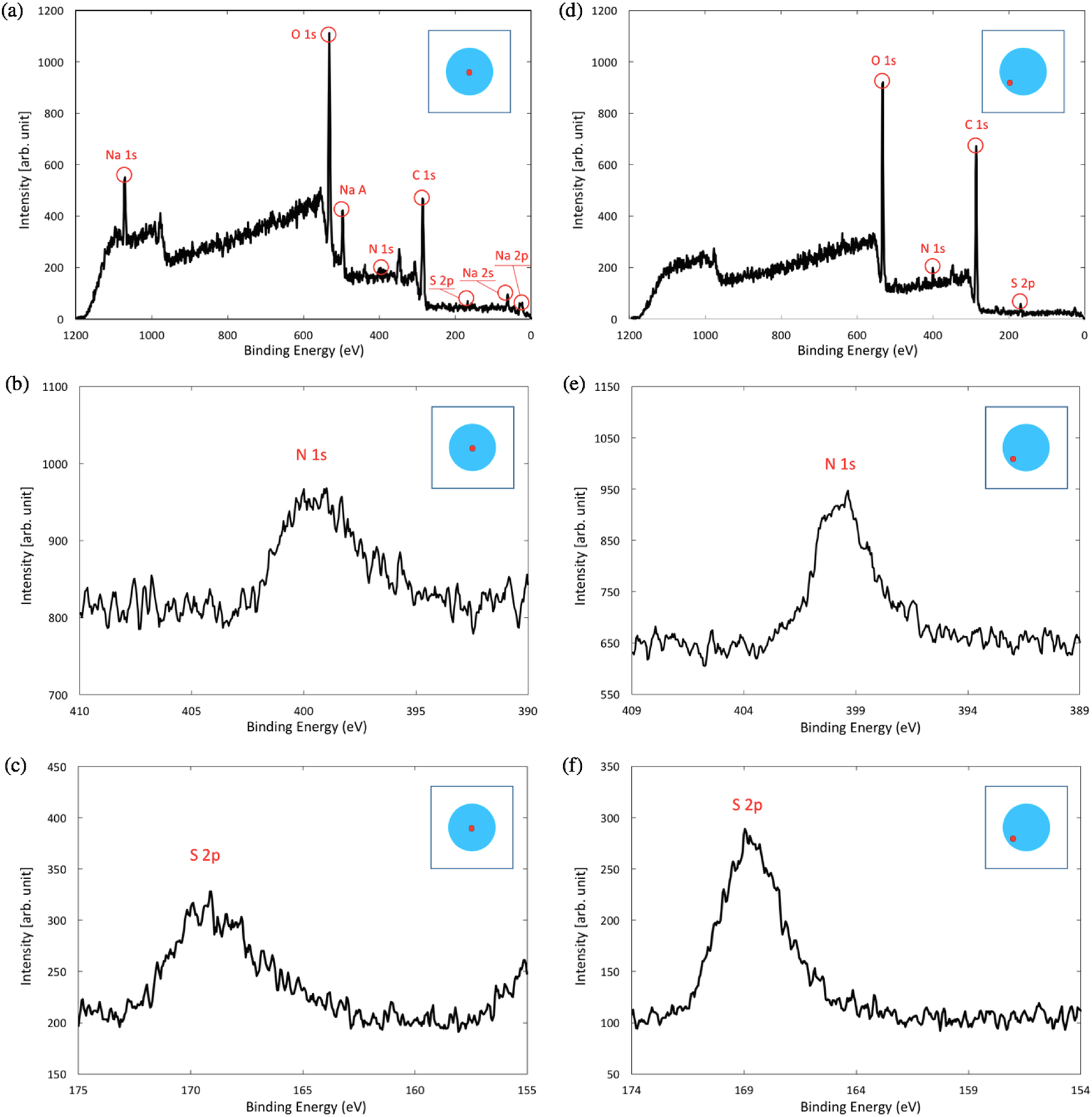}}
\end{minipage}\\%
\flushleft
{\bf Figure S3.} (a-c) XPS spectrum of the central part of the sacran cast film without electrodes.  (b) Result of narrow scan of the binding energy in the 1s orbital of N.  (c) Result of narrow scan of the binding energy in the 2p orbital of S.  (d-f) XPS spectrum of the edge of the sacran cast film without electrodes.  (e) Result of narrow scan of the binding energy in the 1s orbital of N.  (f) Result of narrow scan of the binding energy in the 2p orbital of S.
\label{Figure_S3}
\end{figure*}

\end{document}